\documentstyle[12pt]{article}
\newcommand{\EQ}{\begin{equation}}
\newcommand{\EN}{\end{equation}}

\newcommand{\hs}{\hspace{0.1cm}}

\begin{document}
\oddsidemargin 5mm
\setcounter{page}{0}
\renewcommand{\thefootnote}{\fnsymbol{footnote}}
\newpage
\setcounter{page}{0}
\begin{titlepage}
\begin{flushright}
\hs
\end{flushright}
\vspace{0.5cm}
\begin{center}
{\large {\bf The Sub-leading Magnetic Deformation of
the Tricritical Ising Model in 2D as RSOS Restriction of the
Izergin-Korepin Model}} \\
\vspace{1.5cm}
{\bf F. Colomo$^1$\footnote{Angelo Della Riccia Foundation's
fellow; address after November $15^{th}$, $1991$: I.N.F.N., Sezione di
Firenze, Largo E. Fermi 2, 50125 Firenze, Italy.},
A. Koubek$^2$,
G. Mussardo$^{2}$ \\
\vspace{0.8cm}
$^1${\em NORDITA, Blegdamsvej 17, DK-2100 Copenhagen \O, Denmark }\\
$^2${\em International School for Advanced Studies, Strada Costiera 11,
34014 Trieste, Italy}} \\
\end{center}
\vspace{6mm}
\begin{abstract}
We compute the $S$-matrix of the Tricritical Ising Model perturbed by
the subleading magnetic operator using Smirnov's RSOS reduction of the
Izergin-Korepin model. We discuss some features of the scattering theory
we obtain, in particular a non trivial implementation of crossing-symmetry,
interesting connections between the asymptotic behaviour of the
amplitudes, the possibility of introducing generalized statistics, and the
monodromy properties of the OPE of the unperturbed Conformal Field Theory.
\end{abstract}
\vspace{5mm}
\centerline{published in Phys. Lett. B274 (1992), 367.}
\end{titlepage}

\newpage

\renewcommand{\thefootnote}{\arabic{footnote}}
\setcounter{footnote}{0}

{\bf 1.} Let us consider the quantum field theory defined by perturbing the
fixed-point action of the Tricritical Ising Model (TIM) in 2D by a relevant
operator $\mu(z,\overline z)$ with anomalous dimensions $(\frac{7}{16},
\frac{7}{16})$. This deformation of the massless critical point defines an
integrable massive QFT \cite{Zam,LMC}. The operator $\mu$ is identified as the
field at the position (2,1) in the Kac-table of the model and corresponds to
the scaling limit of the subleading magnetic operator of its lattice version
\cite{BPZ,FQS}. Hence the ${\bf Z}_2$ spin-reversal symmetry of the
tricritical point is explicitly broken and the outcoming massive theory can
exhibit the ``$\Phi^3$-property'' $i.e.$ the absence of a spin $3$ conserved
current and the possibility to have massive particles $A_i$ which appear as
bound states of themselves \cite{Zam}.

The peculiar features of this massive field theory were first underlined
in ref.\hs\cite{LMC} where the model was studied by means of the
``Truncated Conformal Space Approach'' (TCSA) \cite{YZ,LM}.
This approach consists in the diagonalization of the perturbed Hamiltonian
\EQ
H=H_{CFT} + \lambda \int \hs \Phi_{2,1} (x) \hs dx
\EN
on a strip\footnote{We consider in the following only the
case of periodic boundary condition.} of width $R$, truncated at a certain
level of the Hilbert space defined by the conformal field theory at the fixed
point. The lowest energy levels, as functions of $R$, are given in fig.\hs 1.
The spectrum presents two lowest degenerate levels (with an energy splitting
exponentially small for large $R$) and a single bound state $B$ of mass $m$
below the threshold. This is the novelty of the model, which cannot be
explained in terms of a spontaneous symmetry breaking phenomenon. A
spontaneously broken symmetry would imply, in fact, that also the energy
level of the bound state
$B$ should be doubly degenerate, in the large $R$ limit. On the contrary,
as we will show, it is a phenomenon arising from quantum group reduction.
Using a general analysis made by Smirnov \cite{Smirnov}, we compute the exact
factorized $S$-matrix of the model and discuss the interesting features of
this scattering theory.


\vspace{3mm}

{\bf 2.} Smirnov \cite{Smirnov} has proposed a general scheme for constructing
$S$ matrices for $\Phi_{1,2}$ and $\Phi_{2,1}$ perturbations, using
the fact that these theories can be realized as $A_2^{(2)}$ Toda field
theories. He used the general ansatz $S=S_0 R(x,q)$, where
$R$ is the ``R-matrix'' of the affine quantum group $(A_2^{(2)})_q$, and
$S_0$ is a scalar function guaranteeing unitarity. This ansatz does not
work in general but only for special values of the parameter $q$.
For $\Phi_{2,1}$ perturbations of the conformal model $M_{r,r+1}$, the
parameters $x$ and $q$ are related to those of the scattering theory as
$q=e^{2 i \gamma}$ and $x=e^{\frac{2 \pi}{\xi} \beta}$, with $\xi =
\frac{2}{3} \frac{\pi ^2}{2 \gamma - \pi}$, $\gamma = \frac {r}{r+1} \pi $
and $\beta$ is the rapidity variable.
In these cases, q is a root of unity and one can restrict the quantum group
representations such that in tensor-products appear only spins with
$s_i \leq \left[\frac{r-1}2\right]$ ($[x]$ is the integer part
of the number $x$). Transforming to the shadow-world basis \cite{resh88},
which in a statistical language means going to the dual lattice, the $R$
matrix gets essentially replaced by the $6j$ symbols. This amounts in changing
to a new basis for the n-particle scattering states given by
\EQ
\mid \beta_1,j_1, a_1; \hs \beta_2,j_2, a_2; \ldots
\beta_{n-1}, j_{n-1}, a_{n-1}; \hs \beta_n, j_n, a_n\rangle
\EN
wherein $\beta _i$ are the rapidities, $j_i$ are the spins of the particles
($j=0$ for breathers and $j=1$ for kinks) and $a_i$ are numbers, being the
dual variables to the spins, now restricted by the RSOS rules
\EQ
\vert a_k - 1 \vert \, \leq \, a_{k+1} \, \leq \hs
\min \,(a_k + 1, r-2-a_k-j_{k+1}) \hs . \label{limitation}
\EN
For  the TIM perturbed by the subleading magnetization operator, $r=4$,
$a_i=0,1$, and  the one-particle states are the vectors:
$\mid K_{01} \rangle$, $\mid K_{10} \rangle$ and $\mid K_{11} \rangle$.
All of them have the same mass $m$. Notice that, because of the RSOS
restrictions, the state $\mid K_{00}\rangle$ is not allowed. A basis for
the two-particle asymptotic states is
\EQ
\mid K_{01} K_{10} \rangle, \hs\hs
\mid K_{01} K_{11} \rangle, \hs\hs
\mid K_{11} K_{11} \rangle, \hs \hs
\mid K_{11} K_{10} \rangle, \hs \hs
\mid K_{10} K_{01} \rangle \hs .
\label{basis}
\EN
The corresponding scattering processes are
\begin{eqnarray}
\mid K_{01}(\beta_1) K_{10}(\beta_2) \rangle & = &
S_{00}^{11} (\beta_1-\beta_2) \mid K_{01}(\beta_2) K_{10} (\beta_1)
\rangle
\nonumber \\
\mid K_{01}(\beta_1) K_{11}(\beta_2) \rangle & = &
S_{01}^{11} (\beta_1-\beta_2) \mid K_{01}(\beta_2) K_{11} (\beta_1)
\rangle
\nonumber \\
\mid K_{11}(\beta_1) K_{10} (\beta_2)\rangle & = &
S_{10}^{11} (\beta_1-\beta_2) \mid K_{11} (\beta_2) K_{10} (\beta_1)
\rangle
\label{scattering}\\
\mid K_{11} (\beta_1) K_{11} (\beta_2)\rangle & = &
S_{11}^{11} (\beta_1-\beta_2) \mid K_{11} (\beta_2) K_{11} (\beta_1)
\rangle +
S_{11}^{10}(\beta_1-\beta_2) \mid K_{10} (\beta_2) K_{01}
(\beta_1) \rangle \nonumber \\
\mid K_{10} (\beta_1) K_{01} (\beta_2) \rangle & = &
S_{11}^{00} (\beta_1-\beta_2) \mid K_{10} (\beta_2) K_{01} (\beta_1)
\rangle +
S_{11}^{10}(\beta_1-\beta_2) \mid K_{11} (\beta_2) K_{11}(\beta_1)
\rangle \nonumber
\end{eqnarray}
Explicitly, the above amplitudes are given by

\vspace{10mm}

\begin{picture}(190,40)
\thicklines
\put(60,0){\line(1,1){30}}
\put(60,30){\line(1,-1){30}}
\put(60,15){\makebox(0,0){$0$}}
\put(90,15){\makebox(0,0){$0$}}
\put(75,0){\makebox(0,0){$1$}}
\put(75,30){\makebox(0,0){$1$}}
\put(100,15){\makebox(0,0)[l]{$\displaystyle{\hs =\hs S_{00}^{11}(\beta)
\hs =\hs \frac{i}{2}\hs
S_0(\beta)
\hs\sinh\left(\frac{9}{5} \beta -i \frac{\pi}{5}\right)}$}}
\put(390,15){\makebox(0,0)[l]{(6.a)}}
\end{picture}

\vspace{10mm}

\begin{picture}(190,40)
\thicklines
\put(60,0){\line(1,1){30}}
\put(60,30){\line(1,-1){30}}
\put(60,15){\makebox(0,0){$0$}}
\put(90,15){\makebox(0,0){$1$}}
\put(75,0){\makebox(0,0){$1$}}
\put(75,30){\makebox(0,0){$1$}}
\put(100,15){\makebox(0,0)[l]{$\displaystyle{\hs =\hs S_{01}^{11}(\beta)
\hs =\hs -\frac{i}{2}\hs
S_0(\beta)
\hs\sinh\left(\frac{9}{5} \beta +i \frac{\pi}{5}\right)}$}}
\put(390,15){\makebox(0,0)[l]{(6.b)}}
\end{picture}

\vspace{10mm}

\begin{picture}(190,40)
\thicklines
\put(60,0){\line(1,1){30}}
\put(60,30){\line(1,-1){30}}
\put(60,15){\makebox(0,0){$1$}}
\put(90,15){\makebox(0,0){$1$}}
\put(75,0){\makebox(0,0){$1$}}
\put(75,30){\makebox(0,0){$1$}}
\put(100,15){\makebox(0,0)[l]{$\displaystyle{\hs =\hs S_{11}^{11}(\beta)
\hs =\hs \frac{i}{2}\hs S_0(\beta) \hs
\frac{\sin\left(\frac{\pi}{5}\right)}{\sin\left(\frac{2\pi}{5}\right)}
\hs\sinh\left(\frac{9}{5} \beta -i \frac{2\pi}{5}\right)}$}}
\put(390,15){\makebox(0,0)[l]{(6.c)}}
\end{picture}

\vspace{10mm}

\begin{picture}(190,40)
\thicklines
\put(60,0){\line(1,1){30}}
\put(60,30){\line(1,-1){30}}
\put(60,15){\makebox(0,0){$1$}}
\put(90,15){\makebox(0,0){$1$}}
\put(75,0){\makebox(0,0){$0$}}
\put(75,30){\makebox(0,0){$1$}}
\put(100,15){\makebox(0,0)[l]{$\displaystyle{\hs =\hs S_{11}^{01}(\beta)
\hs =\hs -\frac{i}{2}\hs S_0(\beta) \hs
\left(\frac{\sin\left(\frac{\pi}{5}\right)}{\sin\left(\frac{2\pi}{5}\right)}
\right)^{\frac{1}{2}}
\hs\sinh\left(\frac{9}{5} \beta\right)}$}}
\put(390,15){\makebox(0,0)[l]{(6.d)}}
\end{picture}

\vspace{10mm}

\begin{picture}(190,40)
\thicklines
\put(60,0){\line(1,1){30}}
\put(60,30){\line(1,-1){30}}
\put(60,15){\makebox(0,0){$1$}}
\put(90,15){\makebox(0,0){$1$}}
\put(75,0){\makebox(0,0){$0$}}
\put(75,30){\makebox(0,0){$0$}}
\put(100,15){\makebox(0,0)[l]{$\displaystyle{\hs =\hs S_{11}^{00}(\beta)
\hs =\hs -\frac{i}{2}\hs S_0(\beta) \hs
\frac{\sin\left(\frac{\pi}{5}\right)}{\sin\left(\frac{2\pi}{5}\right)}
\hs\sinh\left(\frac{9}{5} \beta +i \frac{2\pi}{5}\right)}$}}
\put(390,15){\makebox(0,0)[l]{(6.e)}}
\end{picture}

\vspace{10mm}

\setcounter{equation}{6}
\noindent

with $S_0$ given by
\begin{eqnarray}
S_0(\beta) &=& -\left(\sinh\frac{9}{10}(\beta-i\pi)
\sinh\frac{9}{10}\left(\beta-\frac{2\pi i}{3}\right)\right)^{-1} \nonumber\\
& & \mbox{} \times w\left(\beta,-\frac{1}{5}\right)
w\left(\beta,+\frac{1}{10}\right)
w\left(\beta,\frac{3}{10}\right) \\
& & \mbox{} \times
t\left(\beta,\frac{2}{9}\right)
t\left(\beta,-\frac{8}{9}\right) t\left(\beta,\frac{7}{9}\right)
t\left(\beta,-\frac{1}{9}\right) \nonumber\hs ,
\end{eqnarray}
where
\[
w(\beta,x)=\frac{\sinh\left(\frac{9}{10} \beta +i\pi x\right)}
{\sinh\left(\frac{9}{10} \beta -i\pi x\right)}\hs ;
\]
\[
t(\beta,x)=\frac{\sinh\hs \frac{1}{2}(\beta +i\pi x)}
{\sinh\hs \frac{1}{2}(\beta -i\pi x)} \hs .
\]

\vspace{5mm}

\noindent
It is easy to check the unitarity relations
\EQ
\sum_{e}
S_{bc}^{ae}(\beta) S_{bc}^{ed}(-\beta) = \delta^{ad} \hs\hs,
\EN
and also the factorization equations
\EQ
\sum_{k} S_{bk}^{ac}(\beta) S_{ae}^{kf}(\beta+\beta^{\prime})
S_{kd}^{ce}(\beta^{\prime})
=\sum_{k} S_{ke}^{df}(\beta) S_{bd}^{ck}(\beta+\beta^{\prime})
S_{ak}^{bf}(\beta^{\prime}) \hs\hs.
\EN
Notice that the crossing symmetry properties occurs in a non trivial way,
$i.e.$
\begin{eqnarray}
S_{11}^{11}(i\pi-\beta) & = & S_{11}^{11}(\beta) \hs ;\nonumber \\
S_{11}^{00}(i\pi-\beta) & = & a^2 \hs S_{00}^{11}(\beta)
\hs ; \label{crossing}\\
S_{11}^{01}(i\pi-\beta) & = & a \hs S_{01}^{11}(\beta) \hs ;\nonumber
\end{eqnarray}
where
\EQ
a=-\left(\frac{s\left(\frac{1}{5}\right)}
{s\left(\frac{2}{5}\right)}\right)^{\frac{1}{2}} \hs ,
\label{asymmetry}
\EN
and $s(x)\equiv \sin(\pi x)$. This is not surprising, since the crossing
operation interchange the relative ordering of kinks to an unphysical one,
and it is necessary to introduce a non trivial charge conjugation operator
to recover a physical ordering (see also \cite{Mussardo}).

The amplitudes (6) are periodic along the imaginary axis of $\beta$
with period $10\hs\pi i$. The structure of poles and zeros is quite
rich. On the physical sheet, $0\leq {\rm Im} \hs\beta\leq i \pi$, the poles
of the $S$-matrix are located at $\beta=\frac{2\pi i}{3} $ and
$\beta=\frac{i\pi}{3}$ (fig.\hs 2). The first pole corresponds to a bound
state in the direct channel while the second one is the singularity
due to the particle exchanged in the crossed process.

It is easy to check that the residue at $\beta=\frac{2\pi i}{3}$ for
the amplitude $S_{00}^{11}(\beta)$ is zero. Hence, in the amplitude
$S_{00}^{11}$ there is no bound state in the direct channel but only the
singularity coming from to the state $\mid K_{11}\rangle$
exchanged in the $t$-channel. This can be seen by stretching
the original amplitudes along the vertical direction
($s$-channel) and along the horizontal one ($t$-channel) (fig.\hs 3).
Since the state $\mid K_{00} \rangle$ is not physical, the residue in the
direct channel turns out to be zero. The same decoupling of the unphysical
state $\mid K_{00} \rangle$ works for the $t$-channel of the amplitude
$S_{11}^{00}$, i.e. in this amplitude the residue of the pole at
$\beta=i\frac{\pi}{3}$ is zero. In all remaining amplitudes, the residues
at the poles $\beta=\frac{2\pi i}{3}$ and $\beta=\frac{i\pi}{3}$ are
different from zero. The corresponding bound states can be identified
with the physical kink states $\mid K_{11}\rangle$, $\mid K_{01} \rangle$ and
$\mid K_{10} \rangle$. For instance, in $S_{11}^{11}$, the state
$\mid K_{11}\rangle$ appears as a bound state in both channels.


\vspace{3mm}

{\bf 3.} The one-particle line $\alpha$ of fig.\hs 1 corresponds to the
state $\mid K_{11}\rangle$. This energy level is not doubly degenerate
because the state $\mid K_{00}\rangle$ is forbidden by the RSOS selection
rules, eq.\hs(\ref{limitation}). With periodic boundary conditions,
the kink states $\mid K_{01}\rangle $ and $\mid K_{10} \rangle $ are
projected out and $\mid K_{11} \rangle $ is the only one-particle state that
can appear in the spectrum. This is the aforementioned explanation of the
spectrum of the subleading magnetic deformation of the TIM.

Let us discuss other properties of the scattering theory under consideration.
For real values of $\beta$, the amplitudes $S_{00}^{11}(\beta)$ and
$S_{01}^{11}(\beta)$ are numbers of modulus 1. It is therefore convenient to
define the following phase shifts
\begin{eqnarray}
S_{00}^{11}(\beta) & \equiv & e^{2 i \delta_{0}(\beta)} \hs;
\label{phaseshift}\\
S_{01}^{11}(\beta) & \equiv & e^{2 i \delta_{1}(\beta)} \hs .\nonumber
\end{eqnarray}
The non-diagonal sector of the scattering processes is characterized
by the $2\times 2$ symmetric $S$-matrix
\EQ
\left(
\begin{array}{ll}
S_{11}^{11}(\beta) & S_{11}^{01}(\beta) \\
S_{11}^{01}(\beta) & S_{11}^{00}(\beta)
\end{array}
\right)
\label{nondiag} \hs .
\EN

\vspace{3mm}
\noindent
We can define the corresponding phase shifts by diagonalizing the matrix
(\ref{nondiag}). The eigenvalues turn out to be the same functions as in
(\ref{phaseshift}), {\em i.e.}
\EQ
\left(
\begin{array}{cc}
e^{2i\delta_0(\beta)} & 0\\
0 & e^{2i\delta_1(\beta)}
\end{array}
\right)
\label{diag} \hs .
\EN
A basis of eigenvectors is given by
\EQ
\mid \phi_i(\beta_1)\phi_i(\beta_2)\rangle \hs = \hs
\sum_{j=0}^1\, U_{ij} \mid K_{1j}(\beta_1) K_{j1}(\beta_2)\rangle
\hs\hs,
\hspace{8mm} i=0,1 \hs\hs,
\EN
where $U$ is an unitary matrix which does not depend on $\beta$
\EQ
U \hs = \hs \frac{1}{\sqrt{1+a^2 }}\hs
\left(
\begin{array}{cc}
1 & a\\
-a &  1
\end{array}
\right)
\label{unitransf} \hs .
\EN
The asymptotic behaviour of the phase shifts is the following:
\begin{eqnarray}
\lim_{\beta\rightarrow\infty} \hs e^{2i \delta_0(\beta)} & = &
e^{\frac{6 \pi i}{5}} \hs ;\\
\lim_{\beta\rightarrow\infty} \hs e^{2 i \delta_1(\beta)} & = &
e^{\frac{3\pi i}{5}}\nonumber \hs .
\label{limits}
\end{eqnarray}
We can use this nontrivial asymptotic values of the phase-shifts in order
to define generalized bilinear commutation relation for
the ``kinks'' $\phi_0$ and $\phi_1$
\cite{Swieca,KT,Smirnov2}
\EQ
\phi_i(t,x) \phi_j (t,y) =\phi_j(t,y) \phi_i(t,x)\hs e^{2\pi i s_{ij}
\epsilon(x-y)} \hs.
\EN
The generalized ``spin'' $s_{ij}$ is a parameter related to the
asymptotic behaviour of the $S$-matrix. A consistent assignment is given by
\begin{eqnarray}
s_{00} & = & \frac{3}{5} =\hs \frac{\delta_0(\infty)}{\pi}\nonumber
\hs\hs;\\
s_{01} & = & 0 \hs\hs;\\
s_{11} & = & \frac{3}{10} =\hs \frac{\delta_1(\infty)}{\pi} \hs\hs.\nonumber
\end{eqnarray}
Notice that the previous monodromy properties are those of the chiral field
$\Psi=\Phi_{\frac{6}{10},0}$ of the original CFT of the TIM.
This field occupies the position $(1,3)$ in the Kac-table of the model.
The operator product expansion of $\Psi$ with itself reads
\EQ
\Psi(z) \Psi(0) = \frac{1}{z^{\frac{6}{5}}} \hs {\bf 1} +
\frac{C_{\Psi,\Psi,\Psi}}{z^{\frac{3}{5}}} \hs \Psi(0) + \ldots
\label{OPE}
\EN
where $C_{\Psi,\Psi,\Psi}$ is the structure constant of the OPE algebra.
Moving $z$ around the origin, $z\rightarrow e^{2 \pi i} z$, the
phase acquired from the first term on the right hand side of (\ref{OPE})
comes from the conformal dimension of the operator $\Psi$ itself. In contrast,
the phase obtained from the second term is due to the insertion of an
additional operator $\Psi$. A similar structure appears in the scattering
processes of the ``kinks'' $\phi_i$: in the amplitude of the kink $\phi_0$
there is no bound state in the s-channel (corresponding to the
``identity term'' in (\ref{OPE})) whereas in the amplitude of $\phi_1$ a kink
can be created as a bound state for $\beta=\frac{2\pi i}{3}$
(corresponding to the ``$\Psi$ term'' in (\ref{OPE})). In the
ultraviolet limit, the fields $\phi_i$ should give rise to the operator
$\Psi(z)$, similarly to the case analyzed in \cite{Smirnov2}. The
actual proof requires the analysis of the form factors and will be given
elsewhere.

The previously discussed fact happens to be a particular case of a general
situation of the RSOS $S$-matrices coming from Smirnov's reduction.
If we study Smirnov's formula \cite{Smirnov} for generic values of
$r\ge 5$, $i.e.$ we consider the $\Phi_{2,1}$ deformation of
${\cal M}_{r,r+1}$, we see that we have
$0\le a_i\le \left[\frac{r-1}{2}\right]$ and therefore we get many more
amplitudes. In general we have three independent diagonal amplitudes,
$S_{00}^{11}(\beta)=e^{2i\delta_0(\beta)}$,
$S_{01}^{11}(\beta)=e^{2i\delta_1(\beta)}$ and
$S_{02}^{11}(\beta)=e^{2i\delta_2(\beta)}$ which define three independent
phase-shifts. Their asymptotic behaviour is
\begin{eqnarray}
\lim_{\beta\rightarrow \infty} \hs S_{00}^{11}(\beta) & = &
e^{2 i \pi \Delta_{1,3} } \hs,\nonumber \\
\lim_{\beta\rightarrow\infty} \hs S_{01}^{11}(\beta) & = &
e^{ i \pi \Delta_{1,3} } \hs ,\label{asympbeh}
\\
\lim_{\beta\rightarrow\infty} \hs S_{00}^{11}(\beta) & = &
e^{ i \pi (2\Delta_{1,3} - \Delta_{1,5}) } \hs,\nonumber
\end{eqnarray}
where $\Delta_{1,3}$ and $\Delta_{1,5}$ are the anomalous dimensions
of the corresponding fields of the original CFT.
The non-diagonal sector of the scattering processes is a block-diagonal
matrix, with $3\times 3$ and $2\times 2 $ non-diagonal blocks.
It may be diagonalized as well, and we get as eigenvalues of all the
$3\times 3$ blocks the functions $e^{2i\delta_0(\beta)}$,
$e^{2i\delta_1(\beta)}$ and $e^{2i\delta_2(\beta)}$, whereas
as eigenvalues of all the $2\times 2$ matrices we get the functions
$e^{2i\delta_0(\beta)}$ and $e^{2i\delta_1(\beta)}$. No other independent
functions appear. Therefore, also the asymptotic behaviour
in the non-diagonal sector is given by (\ref{asympbeh}). In the context of
the previously discussed  monodromy properties, this appears to be related to
the OPE of $\Psi\equiv\Phi_{1,3}$ of the original CFT ${\cal M}_{r,r+1}$:
\EQ
\Psi(z) \Psi(0) = \frac{1}{z^{2\Delta_{1,3}}} \hs {\bf 1} +
\frac{C_{\Psi,\Psi,\Psi}}{z^{\Delta_{1,3}}} \hs \Psi(0) +
\frac{C_{\Psi,\Psi,\Phi_{1,5}}}{z^{2\Delta_{1,3}-\Delta_{1,5}}} \hs
\Phi_{1,5}(0) + \ldots
\EN
In the previous case of TIM, with $r=4$, the last channel could
not be open because $\Phi_{1,5}$ does not appear in the Kac-table of the
primary fields of the original CFT, and the singular part of the OPE
stops after the two first terms. The opening of the new channel $\Phi_{1,5}$
for $r\ge 5$ corresponds to the appearance of states (forbidden by the RSOS
selection rules, eq.\hs(\ref{limitation}) ) with spin $j=2$ in the $s$-channel
of $S_{i-1,i+1}^{i,i}(\beta)$, $i=1,...,\left[\frac{r-1}{2}\right]-1$.

As final remark, we want now to show that is possible to implement the
crossing symmetry for the amplitudes (6) in a standard fashion.
This requires the introduction of the ``gauge'' transformed amplitudes
\cite{Leclair-b}, $i.e$ a change of basis in the space of asymptotic states
\begin{eqnarray}
&\,&\tilde{S}_{a_{k-1} a_{k+1}}^{a_k a_k'}\left(\beta_k-\beta_{k+1}\right)
= \nonumber \\
&=&(-1)^{(a_{k-1}+ a_{k+1}-a_k-a_k')/2}
\left ( \frac{ [ 2 a_k+1]_q[2a_k'+1]_q}{[2a_{k-1} +1]_q[2a_{k+1}+1]_q}
\right )^{-\frac{\beta}{2 \pi i}}
S _{a_{k-1} a_{k+1}}^{a_k a_k'}\left(\beta_k-\beta_{k+1}\right)
\hs,\label{gauge}
\end{eqnarray}
where
\[
[y]_q \hs = \hs \frac{q^{y/2}-q^{-y/2}}{q^{1/2}-q^{-1/2}} \hs\hs.
\]
In the new basis, the crossing relations become trivially
\EQ
\tilde{S} _{a_{k-1} a_{k+1}}^{a_k a_k'}\left(i \pi -\beta \right)
= \tilde{S} _{a_k' a_k}^{a_{k-1} a_{k+1}}\left(\beta\right)
\hs .
\EN
But, the price we pay by performing such procedure is that the new amplitudes
have an oscillatory behaviour for $\beta\rightarrow\infty$, and the
interesting connection with generalized statistics and with the OPE of the
original CFT at the critical point are spoilt.

\vspace{3mm}


{\bf 4.} Following a general framework proposed by Smirnov \cite{Smirnov},
we have constructed the $S$ matrix for the massive theory arising
from the $\Phi_{2,1}$ deformation of TIM. It gives a theoretical foundation
of the observed asymmetry of the spectrum obtained by the Truncation Conformal
Approach \cite{LMC}. We have discussed the phase-shifts and their role
in defining a generalized statistics for the kink excitations.

An alternative scattering theory for the subleading magnetic deformation
of the tricritical Ising model has also been proposed by Zamolodchikov
\cite{Zamsub}. A comparison between the two $S$-matrices (based on
the Truncated Conformal Approach \cite{YZ,LM} and the prediction of finite-size
corrections \cite{Luscher}) is discussed in a separate publication \cite{CKM}.

\section*{Acknowledgments}
We thank T. Miwa, A. Schwimmer and Al.B. Zamolodchikov for useful
discussions. One of us (F.C.) thanks P. Di Vecchia for warm hospitality
at NORDITA.


\newpage
\pagestyle{empty}

\hs

\vspace{25mm}

{\bf Figure Captions}

\vspace{1cm}

\begin{description}
\item[ Figure 1].  First energy levels for the subleading magnetic
perturbation of TIM with periodic boundary conditions on the strip.
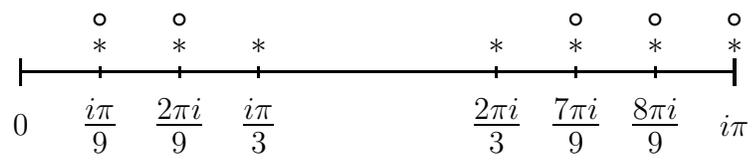
\item[ Figure 2]. Pole structure of $S_0(\beta)$: ${\bf \ast}$ are the
location of the poles and ${\bf o}$ the position of the zeros.
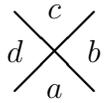
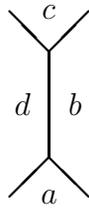
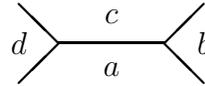
\item[ Figure 3]. Intermediate states in the s-channel and t-channel of
the RSOS $S$-matrix (6).
\end{description}

\newpage

\hs

\begin{center}
\begin{picture}(290,400)
\thicklines
\put(0,60){\line(1,0){270}}
\put(0,55){\line(0,1){10}}
\put(30,58){\line(0,1){4}}
\put(60,58){\line(0,1){4}}
\put(90,58){\line(0,1){4}}
\put(180,58){\line(0,1){4}}
\put(210,58){\line(0,1){4}}
\put(240,58){\line(0,1){4}}
\put(270,55){\line(0,1){10}}
\put(0,40){\makebox(0,0){$0$}}
\put(270,40){\makebox(0,0){$i\pi$}}
\put(30,40){\makebox(0,0){{\Large $\frac{i\pi}{9}$}}}
\put(60,40){\makebox(0,0){{\Large $\frac{2\pi i}{9}$}}}
\put(90,40){\makebox(0,0){{\Large $\frac{i\pi}{3}$}}}
\put(180,40){\makebox(0,0){{\Large $\frac{2\pi i}{3}$}}}
\put(210,40){\makebox(0,0){{\Large $\frac{7\pi i}{9}$}}}
\put(240,40){\makebox(0,0){{\Large $\frac{8\pi i}{9}$}}}
\put(30,70){\makebox(0,0){${\bf \ast}$}}
\put(32,80){\makebox(0,0){\circle{4}}}
\put(60,70){\makebox(0,0){${\bf \ast}$}}
\put(62,80){\makebox(0,0){\circle{4}}}
\put(90,70){\makebox(0,0){${\bf \ast}$}}
\put(180,70){\makebox(0,0){${\bf \ast}$}}
\put(210,70){\makebox(0,0){${\bf \ast}$}}
\put(212,80){\makebox(0,0){\circle{4}}}
\put(240,70){\makebox(0,0){${\bf \ast}$}}
\put(242,80){\makebox(0,0){\circle{4}}}
\put(270,70){\makebox(0,0){${\bf \ast}$}}
\put(272,80){\makebox(0,0){\circle{4}}}
\put(145,0){\makebox(0,0)[t]{{\large Figure 2}}}
\end{picture}
\end{center}

\newpage

\begin{center}
\begin{picture}(450,300)
\thicklines
\put(75,30){\makebox(0,0){${\bf S}$}}
\put(200,30){\makebox(0,0){{\bf s-channel}}}
\put(350,30){\makebox(0,0){{\bf t-channel}}}
\end{picture}

\begin{picture}(450,110)
\thicklines
\put(60,30){\line(1,1){30}}
\put(60,60){\line(1,-1){30}}
\put(60,45){\makebox(0,0){$d$}}
\put(90,45){\makebox(0,0){$b$}}
\put(75,30){\makebox(0,0){$a$}}
\put(75,60){\makebox(0,0){$c$}}
\put(200,25){\line(0,1){40}}
\put(200,25){\line(-1,-1){15}}
\put(200,25){\line(1,-1){15}}
\put(200,65){\line(-1,1){15}}
\put(200,65){\line(1,1){15}}
\put(200,10){\makebox(0,0){$a$}}
\put(200,80){\makebox(0,0){$c$}}
\put(190,45){\makebox(0,0){$d$}}
\put(210,45){\makebox(0,0){$b$}}
\put(330,45){\line(1,0){40}}
\put(330,45){\line(-1,-1){15}}
\put(330,45){\line(-1,1){15}}
\put(370,45){\line(1,1){15}}
\put(370,45){\line(1,-1){15}}
\put(315,45){\makebox(0,0){$d$}}
\put(385,45){\makebox(0,0){$b$}}
\put(350,35){\makebox(0,0){$a$}}
\put(350,55){\makebox(0,0){$c$}}
\end{picture}

\begin{picture}(450,80)
\thicklines
\put(210,30){\makebox(0,0){{\large Figure 3}}}
\end{picture}
\end{center}


\newpage
\begin{thebibliography}{99}
\bibitem{Zam} A.B. Zamolodchikov, {\em JETP Letters} {\bf 46} (1987),160;
{\em Int. Journ. Mod. Phys.} {\bf A3} (1988), 743;
{\em Integrable field theory from CFT}, Proceeding of the Taniguchi Symposium,
Kyoto 1988, to appear in {\em Advanced Studies in Pure Mathematics}.
\bibitem{LMC} M. L\"{a}ssig, G. Mussardo and J.L. Cardy, {\em Nucl. Phys.}
{\bf B348} (1991), 591.
\bibitem{BPZ} A.A. Belavin, A.M. Polyakov and A.B. Zamolodchikov,
{\em Nucl. Phys.} {\bf B241} (1984), 333.
\bibitem{FQS} D. Friedan, Z. Qiu and S. Shenker, {\em Phys. Rev. Lett.}
{\bf 52} (1984), 1575.
\bibitem{YZ} V.P. Yurov and Al.B. Zamolodchikov, {\em Int. J. Mod.
Phys.} {\bf A5} (1990), 3221.
\bibitem{LM} M. L\"{a}ssig and G. Mussardo, {\em Computer Phys. Comm.}
{\bf 66} (1991), 71.
\bibitem{Smirnov}F.A. Smirnov, {\em Int. J. Mod. Phys.} {\bf A6} (1991),
1407.
\bibitem{IK}A.G. Izergin and V.E. Korepin, {\em Commun. Math. Phys.}
{\bf 79} (1981), 303.
\bibitem{resh88} A.N. Kirillov and N. Yu. Reshetikhin, {\em
Representations of the algebra $U_q(sl(2))$, $q$-orthogonal polynomials
and invariants of links}, LOMI-preprint E-9-88;
\bibitem{Mussardo} G. Mussardo, {\em Integrable deformations of the
non-unitary minimal conformal model ${\cal M}_{3,5}$}, NORDITA/91/54,
{\em Int. Jour. Mod. Phys.}{\bf A}, in press.
\bibitem{Swieca} A. Swieca, {\em Fortschr. Phys.} {\bf 25} (1977), 303.
\bibitem{KT} M. Karowski and H.J. Thun, {\em Nucl. Phys.} {\bf B190
[FS3]} (1981), 61.
\bibitem{Smirnov2} F.A. Smirnov, {\em Comm. Math. Phys.} {bf 132}
(1990), 415.
\bibitem{Leclair-b} D. Bernard and A. Leclair, {\em Nucl. Phys.}
{\bf B340} (1990), 721.
\bibitem{Zamsub} A.B. Zamolodchikov, {\em S-matrix of the Subleading
Magnetic Perturbation of the Tricritical Ising Model}, PUTP 1195-90.
\bibitem{Luscher} M. L\"{u}scher, {\em Comm. Math. Phys.} {\bf 104}
(1986), 177; in {\em Progress in Gauge Field Theory} (Carg\`{e}se 1983),
ed. G. 't Hooft et al., Plenum, New York 1984;in {\em Champs, Cordes, et
Ph\'{e}nom\`{e}nes Critiques}, proceedings of the 1988 Les Houches
Summer School, ed. E. Brezin and J. Zinn-Justin, North Holland,
Amsterdam, 1989.
\bibitem{CKM} F. Colomo, A. Koubek and G. Mussardo, {\em On the $S$-matrix of
the Sub-Leading Magnetic Deformation of the Tricritical Ising Model in
Two Dimensions}, NORDITA 91/47, {\em Int. Jour. Mod. Phys.} {\bf A}, in press.
\end{thebibliography}
\end{document}